\documentclass[12pt,a4paper]{article}
\usepackage{amssymb}

\newcommand{\diag}{\mathop{\rm\textstyle{diag}}\nolimits}

\title{
Affine Toda Solitons and Systems of Calogero-Moser Type}

\author{H. W. Braden and Andrew N. W. Hone\\
\normalsize
\em Department of Mathematics and Statistics,\\
\normalsize
\em The University of Edinburgh, \\
\normalsize
\em Edinburgh, UK. \\
\normalsize
e-mail: hwb@ed.ac.uk\quad\quad  hone@maths.ed.ac.uk
\\
}

\begin{document}

\renewcommand{\thepage}{}
\begin{titlepage}
\maketitle

\begin{abstract}
The solitons of affine Toda field theory are related to the
spin-generalised Ruijsenaars-Schneider (or relativistic Calogero-Moser)
models. This provides the sought after extension of the  
correspondence between the sine-Gordon solitons and the 
Ruijsenaars-Schneider model.

\end{abstract}
\vfill
\end{titlepage}
\renewcommand{\thepage}{\arabic{page}}

\section{Introduction}
The purpose of this letter is to relate the solitons of the affine Toda system
to the  spin extensions of the Ruijsenaars-Schneider model,
the latter also being known as the relativistic Calogero-Moser model. 
This work generalises the known connection between the solitons of the sine-Gordon
equation and the (non-spin) Ruijsenaars-Schneider model.
The connection made here should be viewed as part of a larger programme
that seeks to identify  classes of solutions of PDEs with 
finite dimensional mechanical systems, whereby  
the evolution of the solutions to the PDE is expressed as a dynamical
system on the (finite-dimensional) moduli
space of solutions. Thus, for example, the pole solutions of the KP equation
\cite{Krichever, Shiota}
and its reductions 
(such as KdV \cite{Airault}) are related to 
the non-relativistic Calogero-Moser model,  
while the sine-Gordon solitons are related to its relativistic
counterpart, the Ruijsenaars-Schneider model (see \cite{BB, RS}). 
This programme also extends to include the peakon solutions
appearing in fibre-optics and shallow water waves which have associated 
mechanical systems \cite{cal, CH}.
A similar connection may well underlie the appearance of 
finite dimensional mechanical systems in the study of various models
possessing duality \cite{wd, mw}. As the 
dynamics of mechanical systems are often easier
to understand (or simulate) than the equations of motion for a field theory,
such a programme aims at giving qualitative information about field theories
by an appropriate reduction of degrees of freedom.
The recent work of Babelon, Bernard and Smirnov \cite{BBS}  
may be viewed as taking this correspondence between 
field theories and mechanical systems beyond the classical to include
the quantum regime as well, though the ability to focus attention solely on a
fixed
N-particle sector of the full quantum Hilbert space appears to depend
crucially on the model. Our work will reveal further new features
in such correspondences, as well as provide a sought after generalisation
of known results about the sine-Gordon model to 
the case of affine Toda solitons.

Ruijsenaars and Schneider's seminal work \cite{RS} showed that the soliton
solutions of a variety of equations were related to dynamics built
from the Hamiltonians (with canonically conjugate variables $q_j$, $p_j$)
$$
H_\pm = \sum_j\sp{N} e\sp{\pm p_j}\ \prod\sp{N}_{k\not= j}
\coth\biggl({\frac{q_j-q_k}{2} }\biggr) ,
$$
and equations of motion (for either $H_\pm$)
\begin{equation}
\ddot q_{j}=2\sum_{k\neq j}\frac{\dot q_{j}\dot q_{k}}{\sinh(q_{j}-q_{k})}.
\label{eq:rseqm}
\end{equation}
An appropriate scaling limit of $H=H_+ + H_-$ yields a system of
Calogero-Moser type. 
In particular, the eigenvalues $i\,e\sp{q_{j}}$ of an $N\times N$
matrix associated with the tau function describing an N-soliton solution
of the sine-Gordon equation evolve according to (\ref{eq:rseqm}).
The $q_{j}$'s and $p_{j}$'s may, at least when they are well separated,
be related to the positions and rapidities of N constituent single-solitons;
the dynamics of the system encodes the various soliton phase shifts.
(More details of this will be given below.)
Thus the system governed by $H_+$ describes how the space-time
trajectories of the \lq constituent\rq\ solitons interact.
Of course the same system may be described via the inverse
scattering transform by a free system with linearly evolving data: the
point of the Ruijsenaars-Schneider description is to make greater contact
with the particle description of the soliton. 

Viewing the sine-Gordon model as the $A\sp{(1)}_{1}$ affine Toda system
(with imaginary coupling) a natural question to ask is how the above results
generalise to other affine Toda systems. These systems have been
extensively studied in recent years both for real and imaginary couplings.  
In the real coupling regime a beautiful structure  was uncovered and
exact S-matrices have been conjectured for the theories (see 
\cite{BCDS, CM, Dor} and references therein).
The imaginary coupling regime has also been investigated and classically the
solitons have real energy-momentum although the Lagrangian is complex
(\cite{otu} and references therein).
Spence and Underwood \cite{spu} 
have recently used this work to obtain the symplectic
form on the space of affine Toda solitons but a dynamical description
generalising the sine-Gordon/Ruijsenaars-Schneider
correspondence has proved elusive. 
The purpose of the present letter is to give this
generalisation. Just as the affine Toda systems generalise the sine-Gordon
model, there are spin-generalisations of the Ruijsenaars-Schneider
systems, and it is these
systems  which describe the  dynamics of the affine Toda solitons.
These models (which have been most studied in the $A_n$ setting) are the
relativistic extension of Gibbon and Hermsen's spin generalisation of
the original Calogero-Moser model \cite{GH}.
One new feature we have found in our correspondence is the
appearance of new degrees of freedom, the internal spins of the
model. Although not needed to describe the solitons of the
affine Toda system, these spins determine the matrix that diagonalises
the Lax pair. We will comment further on this later in the letter.

An outline of the letter is as follows. First we will review the
construction of affine Toda solitons, and then in section 3 consider the
reduced symplectic form of the theory. We are then in a position to
relate the affine Toda solitons to the spin-generalised
Ruijsenaars-Schneider model in section 4.
For the purposes of this letter we shall 
limit our discussion to the $A\sp{(1)}_n$ case,
both for simplicity and to make clear the
generalisation of the sine-Gordon/Ruijsenaars-Schneider correspondence.

\section{The $A_{n}\sp{(1)}$ Affine Toda Solitons} 

For the $A_{n}\sp{(1)}$ affine Toda theory with imaginary coupling, 
the equations of motion read 
 
\begin{equation}
\partial_{+}\partial_{-}\phi_{j} +\frac{m\sp{2}}{2i\beta}
(e\sp{i\beta(\phi_{j}-\phi_{j+1})}-e\sp{i\beta(\phi_{j-1}-\phi_{j})})=0, 
\label{eq:mot} 
\end{equation} 
$j=0,1,...,n$.
Here $\pm$ denotes differentiation with respect to light-cone coordinates 
$x_{\pm}=\frac{1}{\sqrt2}(t \pm x)$, and the indices on the components of 
the field $\phi$ are read modulo $(n+1)$ where necessary. 
We shall be considering the solitonic sector of the theory, which means 
assuming $\sum_{j=0}\sp{n}\phi_{j}=0$
(in other words, discarding the free 
field part of $\phi$). 

There are various ways to construct and parametrise soliton 
solutions to (\ref{eq:mot}). Perhaps the simplest 
methods to implement from a 
practical point of view are the application of 
the B\"{a}cklund transformation derived by Fordy and Gibbons \cite{FG}   
or the bilinear formalism developed by Hirota \cite{hir}.  
There are also the powerful vertex operator techniques which make full use of 
the representation theory of the $A\sp{(1)}_{n}$ algebra \cite{otu}. 
While the latter approach is currently the most popular, 
we wish 
to make contact with the original work of Ruijsenaars and Schneider 
\cite{RS}, 
which made much reference to the soliton formulae 
of Hirota. Hence we choose to  
start from the form of the N-soliton solution 
of (\ref{eq:mot}) derived by Hollowood \cite{hol} via 
Hirota's direct method.   
The $l$-th component of the field $\phi$ is given by 
\begin{equation}
e\sp{i\beta\phi_{l}}=\frac{\tau_{l-1}}{\tau_{l}},
\end{equation}
where the tau function $\tau_{l}$ is of the form 
\begin{equation}
\tau_{l}=\sum_{\epsilon}\exp
\left(\sum_{j<k}
\epsilon_{j}\epsilon_{k}B_{jk}+
\sum_{j}\epsilon_{j}\zeta_{j,l}(x_{+},x_{-})\right).
\label{eq:taufn}
\end{equation}
In the above the $\epsilon$ indicates a summation over all possible
combinations of $\epsilon_{j}$ taking the values $0$ or $1$, and the 
indices $j$ and $k$ take values in $\{1,...,N\}$. 
We will explain shortly what the various terms in (\ref{eq:taufn}) mean, and 
how we have parametrised the $A_{n}\sp{(1)}$ affine Toda solitons.  
For the moment we would like to comment that  expression (\ref{eq:taufn}) 
is a rather generic form of the soliton tau function for an integrable PDE,   
the precise nature of $B_{jk}$ and $\zeta_{j,l}$ depending on the 
particular PDE being considered; it may be viewed as a degeneration
of the theta function solutions of the PDE given via algebraic geometry in
which the $\epsilon_{j}$'s run over all of the integers.
Ruijsenaars and Schneider succeeded in making 
the connection between their relativistic Calogero-Moser systems and soliton 
solutions of the sine-Gordon and KdV equations, among others, by showing a 
direct correspondence between the coordinates of the N-particle system and 
the parameters of the N-soliton solution. 
An important part of the correspondence was that all the tau functions 
of form (\ref{eq:taufn}) being considered in \cite{RS} could 
be written in terms of determinants like  
\[ 
\det\,(1+M) 
\] 
for suitable matrices $M$. In what follows we express all the $N$-soliton 
solutions of the $A\sp{(1)}_{n}$
affine Toda theory in this way, and thereby  
obtain a relation to spin-generalised Ruijsenaars-Schneider systems.

First of all we should explain the parameters of the 
Toda N-soliton which appear in 
(\ref{eq:taufn}). Each soliton has a rapidity denoted by $\eta_{j}$, 
a position parameter denoted by $a_{j}$,
and a discrete parameter $\theta_{j}$ taking values in
$\{\frac{2\pi k}{n+1}|k=1,2,...,n\}$ (so that $\exp(i\theta_{j})$ is an
$(n+1)$th root of unity).  
The rapidities are all real, while the $a_{j}$ are pure imaginary for solitons 
(there are different reality conditions for other types of solution e.g. 
breathers). The different values of $\theta_{j}$ give $n$ different species 
of soliton in the $A_{n}\sp{(1)}$ affine Toda theory whose masses are
$2 m \sin(\theta_{j}/2)$.  We also need to define 
\[
\mu\sp{\pm}_{j}=\exp(\eta_{j}\pm \frac{1}{2}i\theta_{j}).
\] 
With this choice of parameters, the terms in the sum (\ref{eq:taufn}) are 
given by 
\[
B_{jk}=\log\left(
\frac{(\mu_{j}\sp{+}-\mu_{k}\sp{+})(\mu_{j}\sp{-}-\mu_{k}\sp{-})}
{(\mu_{j}\sp{-}-\mu_{k}\sp{+})(\mu_{j}\sp{+}-\mu_{k}\sp{-})}
\right),
\]
\[
\zeta_{j,l}(x_{+},x_{-})=\log\left(
a_{j}\exp\left(\sqrt{2}m(e\sp{-\eta_{j}}x_{+}-e\sp{+\eta_{j}}x_{-})
\sin(\theta_{j}/2)+il\theta_{j}\right) 
\right).
\] 
(To make a comparison with the vertex operator formulae, we note that in terms 
of the notation of reference 
\cite{spu}, we have $B_{jk}=\log(X_{j,k})$, 
$a_{j}=Q_{j}$. We will deal with the general formalism elsewhere.)  

We are now ready to write the tau functions as determinants. In fact Olive, 
Turok and Liao \cite{lot} 
found that determinants naturally arose when they derived 
the N-soliton solution by the B\"{a}cklund transformation, but the 
matrices involved are not of the right form for our purposes. Instead we set 
$X_{j}=a_{j}(\mu_{j}\sp{+}-\mu_{j}\sp{-})
\exp\left(
\sqrt{2}m(e\sp{-\eta_{j}}x_{+}-e\sp{\eta_{j}}x_{-})\sin(\theta_{j}/2)
\right)$,
and define $N$-by-$N$ matrices $V$, $\Theta$ by
\begin{equation}
V_{jk}=\frac{\sqrt{X_{j}X_{k}}}{\mu_{j}\sp{+}-\mu_{k}\sp{-}}, 
\label{eq:vdef}
\end{equation}
and
\[
\Theta=\diag(\theta_{1},\theta_{2},...,\theta_{N}).
\]
Then we find that
\begin{equation}
\tau_{l}=\det\hspace{0.05in}(1+e\sp{il\Theta/2}Ve\sp{il\Theta/2}).
\label{eq:tauv}
\end{equation}
To verify (\ref{eq:tauv}) it is necessary to expand the determinant on the
right-hand side in terms of the principal cofactors of $V$, and then use
Cauchy's identity:
\[
\det\left( \frac{1}{\mu_{j}\sp{+}-\mu_{k}\sp{-}}\right)_{j,k}=
\prod_{j}\frac{1}{\mu_{j}\sp{+}-\mu_{j}\sp{-}}
\prod_{j<k}\frac{(\mu_{j}\sp{+}-\mu_{k}\sp{+})(\mu_{j}\sp{-}-
\mu_{k}\sp{-})}
{(\mu_{j}\sp{+}-\mu_{k}\sp{-})(\mu_{j}\sp{-}-\mu_{k}\sp{+})} .
\]
Writing the right-hand side of (\ref{eq:taufn}) in terms of these new
parameters and comparing with the cofactor expansion gives the
required result. Note that in the $A_{1}\sp{(1)}$ (sine-Gordon) case
the $\theta_{j}$ must all take the value $\pi$, which means that
the matrix exponentials appearing in (\ref{eq:tauv}) are multiples of
the identity, and we reproduce the standard result
\[
e\sp{i\beta\phi_{0}}=e\sp{-i\beta\phi_{1}}=\frac{\det\hspace{0.05in}(1-V)}
{\det\hspace{0.05in}(1+V)}.
\]

\section{The Reduced Symplectic Form} 

In this section we describe the phase space of the $N$-soliton solution in 
terms of its symplectic form, before describing how spin-generalised 
Ruijsenaars-Schneider systems arise in the following section.   
The phase space of the affine Toda system has the standard symplectic
form
\begin{equation}
\Omega=\int_{-\infty}\sp{\infty}(\delta \phi_{t}\wedge_{,}\delta \phi)
\hspace{.05in}dx.
\label{eq:symp}
\end{equation}
On substitution of
the $N$-soliton solution into
(\ref{eq:symp}), one obtains (after an integration)
the reduced symplectic form on the
$N$-soliton phase space. In practice it is not possible to perform the
integration for anything other than the one-soliton solution 
\cite{spu}  
(except for the sine-Gordon case, where Babelon and Bernard 
succeeded
in showing that the integrand could be written as an exact derivative
for both the one- and two-soliton 
\cite{BB}).
For the one-soliton phase space, the reduced symplectic form is
(up to an irrelevant numerical factor independent of $\theta$)
\[
\omega\sp{(1)}=\frac{da}{a}\wedge d\eta.
\]

The intractability of the integral (\ref{eq:symp}) for the general
$N$-soliton solution does not matter, as it is a standard result
that as $t\rightarrow\pm\infty$ (the $out/in$ limits) the $N$-soliton
decomposes into a superposition of $N$ one solitons with a shift of
the parameters. So the symplectic form may just be written
\begin{equation}
\omega\sp{(N)}=\sum_{j}\frac{da_{j}\sp{out}}{a_{j}\sp{out}}
\wedge d\eta_{j}\sp{out}
=\sum_{j}\frac{da_{j}\sp{in}}{a_{j}\sp{in}}
\wedge d\eta_{j}\sp{in}.
\label{eq:sols}
\end{equation} 
By direct calculation using the formula (\ref{eq:taufn}) for the tau functions 
of the $N$-soliton solution, we find the relations 
between the $out/in$ parameters and the standard ones: 
\[ 
\eta\sp{in}_{j}=\eta\sp{out}_{j}=\eta_{j}, 
\]   
\[ 
a_{j}\sp{in}=a_{j}  
\prod_{k>j}  
\frac{(\mu_{j}\sp{+}-\mu_{k}\sp{+})(\mu_{j}\sp{-}-\mu_{k}\sp{-})}
{(\mu_{j}\sp{-}-\mu_{k}\sp{+})(\mu_{j}\sp{+}-\mu_{k}\sp{-})}
=a_{j}\prod_{k>j}\exp(B_{jk})   
\]  
(and similarly for $a_{j}\sp{out}$ with the inequality reversed). This agrees 
with the formulae of Spence and Underwood \cite{spu} 
obtained via vertex operator 
arguments, where in their notation $a_{j}=Q_{j}$ and $\exp(B_{jk})=X_{j,k}$.   
So substituting for the $in$ parameters into (\ref{eq:sols}), we obtain the 
N-soliton symplectic form as  
\begin{equation}   
\omega\sp{(N)}=\sum_{j}\frac{da_{j}}{a_{j}}\wedge d\eta_{j} + 
\sum_{j<k}E_{jk}(\eta)\sinh(\eta_{j}-\eta_{k}) 
d\eta_{j}\wedge d\eta_{k}, \label{eq:rsymp}  
\end{equation}  
where 
\[  
E_{jk}(\eta)= 
\frac{1}
{\cosh(\eta_{j}-\eta_{k})-\cos((\theta_{j}-\theta_{k})/2)} 
-\frac{1}
{\cosh(\eta_{j}-\eta_{k})-\cos((\theta_{j}+\theta_{k})/2)}.  
\]  
We observe that $\omega\sp{(N)}$ is clearly real if we choose 
the $\eta_{j}$ to be real and the $a_{j}$ to be pure 
imaginary (which in the $A_{1}\sp{(1)}$ case coincides with the 
condition on $a_{j}$ for sine-Gordon solitons given in \cite{BB}).  
This means that the matrix $V$ defined in (\ref{eq:vdef}) is 
anti-hermitian, which will be important in the next section when we look 
at the dynamics of the eigenvalues of $V$.

\section{Ruijsenaars-Schneider Systems} 

Here we consider how the eigenvalues of the matrix $V$ evolve with 
respect to each of 
the light cone coordinates, and 
find that spin-generalised Ruijsenaars-Schneider equations result.  
Since $V$ is anti-hermitian, it may be diagonalised with a unitary matrix $U$:
\[ 
Q:=UVU\sp{\dagger}=\diag(i\exp(q_{1}),...,i\exp(q_{N})).        
\] 
If we let a dot denote $\frac{d}{dx_{\pm}}$, 
then $V$ satisfies 
\begin{equation}  
\dot V=\frac{1}{2}(\Lambda V+V \Lambda), \label{eq:dot}    
\end{equation}   
for the constant diagonal matrix $\Lambda$ given by 
\[ 
\Lambda=\diag(\pm\sqrt{2}m\exp(\mp\eta_{1})\sin(\theta_{1}/2), 
...,\pm\sqrt{2}m\exp(\mp\eta_{N})\sin(\theta_{N}/2)). 
\] 
Now let $u_{j}$ denote the $j$th row of $U$ (considered as a column vector, 
so that the $u_{j}$ are the left eigenvectors of $V$).
Define the Lax matrix $L$ by
\[
L_{jk}=u_{k}\sp{\dagger}\Lambda u_{j},
\]
so that 
\[ 
L=U\Lambda U\sp{\dagger}. 
\]  
Then $L$ satisfies the Lax equation
\begin{equation}
\dot L=[M,L],  \label{eq:lax}
\end{equation}
for $M=\dot U U\sp{\dagger}$. 
Differentiating the definition of $Q$ gives 
\[ 
\dot Q=[M,Q]+U\dot V U\sp{\dagger}, 
\] 
and after substituting for $\dot V$ from (\ref{eq:dot}) and using the 
definition of $L$ we find  
the identity 
\begin{equation}  
LQ+QL=2(\dot Q+[Q,M]). \label{eq:id}   
\end{equation}  
Upon setting $Q_{j}=i\, e\sp{q_{j}}$,
then (\ref{eq:id}) in components reads
\[ 
L_{jk}(Q_{j}+Q_{k})=2(\dot Q_{j}\delta_{jk} + M_{jk}(Q_{j}-Q_{k})), 
\] 
which yields 
\begin{equation}   
L_{jj}=\dot q_{j} \label{eq:const}  
\end{equation}   
and 
\[ 
M_{jk}=\frac{1}{2}\left(\frac{Q_{j}+Q_{k}} {Q_{j}-Q_{k}} 
\right)L_{jk}=\frac{1}{2}\coth((q_{j}-q_{k})/2)L_{jk}, 
\]
(for $j\neq k$). When $V$ is diagonalised  
we may always choose the phases of the left eigenvectors 
$u_{j}$ so that $M_{jj}=u_{j}\sp{\dagger}\dot u_{j}=0$.  
Substituting these into the Lax equation (\ref{eq:lax}) produces the 
equations of motion: 
\begin{equation}  
\dot L_{jj}=\ddot q_{j}=\sum_{k\neq j}\coth((q_{j}-q_{k})/2)L_{jk}L_{kj}, 
\label{eq:rsma} 
\end{equation}   
\begin{equation}   
\begin{array}{l}
\dot L_{jk}=
\frac{1}{2}\coth((q_{j}-q_{k})/2)(\dot q_{k}-\dot q_{j})L_{jk}\\
\quad\quad\quad+\sum_{l\neq j,k}\frac{1}{2} 
(\coth((q_{j}-q_{l})/2)-\coth((q_{l}-q_{k})/2))
L_{jl}L_{lk}
\end{array}
\label{eq:rsmb} 
\end{equation}  
($j\neq k$).  
These are in fact the spin-generalised Ruijsenaars-Schneider equations with 
certain constraints, although to see this requires comparison with the 
formulae of Krichever and Zabrodin \cite{krz}.  

In \cite{krz} the generalised 
Ruijsenaars-Schneider model is defined in terms of 
$N$ particle positions $x_{j}$ and their internal degrees of freedom (spins)  
given by $l$-dimensional vectors $a_{j}$ and $l$-dimensional covectors 
$b_{j}\sp{\dagger}$, subject to the equations of motion 
\begin{equation} 
\ddot x_{j}=\sum_{k\neq j}(b_{j}\sp{\dagger}a_{k})(b_{k}\sp{\dagger}a_{j}) 
({\cal V}(x_{j}-x_{k})-{\cal V}(x_{k}-x_{j})), 
\label{eq:ra} 
\end{equation}        
\begin{equation}
\dot a_{j}=\sum_{k\neq j}a_{k}(b_{k}\sp{\dagger}a_{j}){\cal V}(x_{j}-x_{k}),  
\label{eq:rb} 
\end{equation} 
\begin{equation} 
\dot b_{j}\sp{\dagger}=-\sum_{k\neq j}b_{k}\sp{\dagger}
(b_{j}\sp{\dagger}a_{k}){\cal V}(x_{k}-x_{j}).
\label{eq:rc} 
\end{equation}
The potential ${\cal V}$ is expressed in terms of the 
Weierstrass zeta function or 
its rational or hyperbolic limits. To make contact with our equations 
we set $x_{j}=q_{j}$ and choose the hyperbolic potential   
\[    
{\cal V}(q_{j}-q_{k})=\frac{1}{2}\coth((q_{j}-q_{k})/2). 
\]  
Then (\ref{eq:ra}) generalises (\ref{eq:rseqm}).
In \cite{krz} the spin degrees of 
freedom were real, but here we allow them to be 
complex, and identify them with the eigenvectors of $V$ by setting 
\[ 
b_{j}\sp{\dagger}=u_{j}\sp{\dagger}, \quad
a_{j}=\Lambda u_{j}. 
\]  
So we have taken $l=N$, and in fact 
our spins are expressed entirely in terms of the eigenvectors of 
$V$ and the constant matrix $\Lambda$; in particular the $b_{j}\sp{\dagger}$ 
must form an orthonormal basis. 
In the notation of \cite{krz}  
the components of the Lax matrix are given by 
\[ 
L_{jk}=b_{k}\sp{\dagger}a_{j}. 
\]  
There are various other constraints that 
we have imposed on our system. First the equations (\ref{eq:ra}),   
(\ref{eq:rb}) and (\ref{eq:rc}) have the scaling symmetry 
\[ 
a_{j}\rightarrow \alpha_{j}a_{j}, \quad
b_{j}\sp{\dagger}\rightarrow \frac{1}{\alpha_{j}}b_{j}\sp{\dagger}. 
\] 
The corresponding integrals of motion are $\dot x_{j}-b_{j}\sp{\dagger}a_{j}$, 
and setting them to zero and rewriting them in terms of 
our coordinates shows that this is 
equivalent to equation (\ref{eq:const}). Similarly our 
requirement that $M_{jj}=0$ is another constraint on the system. 
Now given these constraints we find that from the definition of $L$ 
in terms of the spins we can compute  
$\dot L_{jk}$. So for $j=k$ (\ref{eq:ra}) is equivalent to 
(\ref{eq:rsma}), while for $j\neq k$ (\ref{eq:rb}) and (\ref{eq:rc}) yield 
(\ref{eq:rsmb}).

To make the correspondence between the solitons and the many-body system 
clearer, it is worth considering the sine-Gordon case in more detail and 
comparing it with the general situation. The results about sine-Gordon  
solitons are explained in detail in \cite{BB}, and we have kept our notation 
as similar to this reference as possible to make comparison easier. The 
first thing to observe is that in the $A_{1}\sp{(1)}$ case only knowledge of 
the $q_{j}$ is required to specify the field components, as we have 
\[ 
e\sp{i\beta\phi_{0}}=e\sp{-i\beta\phi_{1}}=
\prod_{j=1}\sp{N}\left( 
\frac{1-i\exp(q_{j})}
{1+i\exp(q_{j})}
\right).
\]
In the general case the presence of the matrix $e\sp{il\Theta/2}$ in 
the expression for the tau functions (\ref{eq:tauv}) means that knowledge 
of both the spin vectors $u_{j}$ (which make up the matrix $U$) and the
$q_{j}$ is required to evaluate these determinants. 
The essential difference is that for sine-Gordon 
there is only one soliton species, while in the $A\sp{(1)}_{n}$ case there 
are $n$ different species corresponding to the different allowed values 
of $\theta_{j}$. This difference is also apparent at the level of the 
equations of motion. In fact when we differentiate the matrix $V$, in the 
case of sine-Gordon we find from (\ref{eq:dot}) that 
\[ 
\dot V=i(ee\sp{\dagger}) 
\] 
for a certain vector $e$. But then conjugating the equation (\ref{eq:dot}) 
with $U$ we obtain 
\[ 
i\tilde{e}\tilde{e}\sp{\dagger}=\frac{1}{2}(LQ+QL), 
\] 
where $\tilde{e}=Ue$. Actually $\tilde{e}$ is a real vector, and in 
terms of its components $\tilde{e}_{j}$  
we have 
\[ 
L_{jk}=2 \,\frac{\tilde{e}_{j}\tilde{e}_{k}}{\exp(q_{j})+\exp(q_{k})}. 
\] 
Since we know the diagonal elements of $L$ explicitly in terms of the 
$q_{j}$ (from (\ref{eq:const})) the above formula means that we then 
know all the $\tilde{e}_{j}$ and hence the off-diagonal elements of $L$ 
are found to be 
\[ 
L_{jk}=\frac{\sqrt{\dot q_{j} \dot q_{k}}}{\cosh((q_{j}-q_{k})/2)}. 
\] 
This may then be substituted into (\ref{eq:rsma}),(\ref{eq:rsmb}) to give the 
ordinary (non-spin) Ruijsenaars-Schneider equations. In this case
(\ref{eq:rsma}) yields (\ref{eq:rseqm}) and (\ref{eq:rsmb})
is a consequence. Babelon and Bernard 
have shown \cite{BB} that there is a canonical transformation 
between the soliton parameters 
and the dynamical variables $q_{j},\dot q_{j}$ (more precisely, 
they formulate 
this in terms of the variables 
$Q_{j}=i\hspace{.05in}\exp(q_{j})$). We discuss 
how this could possibly be extended to the $A_{n}\sp{(1)}$ case in our 
Conclusion. 

\section{Conclusion} 

We have shown the connection between spin-generalised Ruijsenaars-Schneider 
systems and $A_{n}\sp{(1)}$ affine Toda solitons. The soliton 
tau functions are determined by the positions $q_{j}$ of $n$ particles on 
the line as well as an orthonormal set of $n$-dimensional 
spin vectors $u_{j}$, which are together subject to the equations of a 
constrained spin-generalised Ruijsenaars-Schneider model. This extends the 
known result for the sine-Gordon equation, where the spins are no longer 
part of the dynamics and there is a canonical transformation 
between the positions and momenta of the particles and the parameters of 
the solitons. For the general case such a transformation is no longer 
apparent, although we note that the $N$-soliton phase space is still of 
dimension $2 N$, and so it is worth exploring exactly how the 
extra spin degrees 
of freedom are absorbed in the transition from the dynamical variables 
to the soliton parameters. Also it would be interesting to see 
what r\^{o}le the spins might play in the quantum theory. 
Finally there remains the extension to the other affine algebras
and elucidating the connections to the vertex operator constructions
mentioned at various points in the text.
We intend to pursue these points in the future. 

\section{Acknowledgements}
One of us (ANWH) thanks the EPSRC for support.
HWB thanks D. Bernard, D Olive and R Sasaki for stimulating
discussions on related matters.

\end{document}